\documentclass[12pt,a4paper]{article}

\usepackage{a4wide}
\usepackage{amsmath}
\usepackage{bm}
\usepackage{amssymb}
\usepackage{hyperref}
\usepackage{epic}

\newcommand{\Op}{\mathcal{O}}
\newcommand{\HS}{S}
\newcommand{\HBS}{\mathbb S}
\newcommand{\M}{j}

\newcommand{\zt}{\zeta_3}

\newcommand{\zf}{\zeta_5}

\newcommand{\cN}{{\cal N}}

\begin{document}

\thispagestyle{empty}


\begin{center}
{\Large{\bf
The non-planar contribution 
to the four-loop anomalous dimension of twist-2 operators:\\[3mm]
first moments in $\cN=4$ SYM and non-singlet QCD
}}
\vspace{15mm}

{\sc
V.~N.~Velizhanin}\\[5mm]

{\it Theoretical Physics Department\\
Petersburg Nuclear Physics Institute\\
Orlova Roscha, Gatchina\\
188300 St.~Petersburg, Russia}\\[5mm]

\textbf{Abstract}\\[2mm]
\end{center}

\noindent{
We present the result of a \textit{full direct} component calculation for the first three even moments of the non-planar contribution into the four-loop anomalous dimension of twist-2 operators in maximally extended $\cN =4$ supersymmetric Yang-Mills theory.
Obtained result completes our previous calculations in arXiv:0902.4646 and gives the usual result for the higher moments on the contrary to degenerate one in the case of Konishi.
We propose a general form of $\zf$ and $\zt$ parts of the complete non-planar four-loop anomalous dimension of twist-2 operators.
As by product, we have obtained the first even moment of the non-planar contribution to the non-singlet four-loop anomalous dimension of Wilson twist-2 operators in QCD.
}
\newpage

\setcounter{page}{1}

Calculation of anomalous dimensions of the Wilson twist-2 operators is one of the part of operator product expansion for the structure functions
in the framework of perturbative Quantum Chromodynamics (QCD).
At the present time such calculations are performed up to three-loop order~\cite{Gross:1973ju}, while other part of operator product expansion, the coefficient functions, which are known in the same order~\cite{Vermaseren:2005qc}, demand the four-loop anomalous dimensions.

Moreover, the great interest in the calculations of anomalous dimensions of the composite operators comes from the investigations of integrability in the framework of AdS/CFT-correspondence~\cite{Maldacena:1997re}.
In the planar limit there are different calculations~\cite{Anselmi:1998ms,Bajnok:2008bm} for the test of Asymptotic Bethe Ansatz~\cite{Minahan:2002ve} as well as for recently proposed spectral equations for the finite length operators~\cite{Arutyunov:2009zu}.
The calculations are performed perturbatively up to four/five loops~\cite{Anselmi:1998ms} for twist-2/twist-3 operators and up to one loop more with the using of generalized L\"usher corrections~\cite{Luscher:1985dn} for the operators with arbitrary Lorentz spin~\cite{Bajnok:2008bm}.

For non-planar case we calculated some time ago the anomalous dimension of the Konishi operator at fourth order in $\cN=4$ SYM theory~\cite{Velizhanin:2009gv}, where non-planar contribution appears for the first time for the twist-2 operators.
The result was rather surprising, since it contains only $\zf$ contribution without $\zt$ and rational parts:
\begin{equation}
\gamma^{\mathrm {4-loop,\,np}}_{\mathrm{Konishi}}=
-\frac{17280}{N_c^2}\ \zf\,g^8,\qquad g^2\ =\ \frac{g^2_{YM}\,N_c}{(4\,\pi)^2}\,.
\label{resnpK}
\end{equation}
The calculations of the $\zf$ contribution to the first  three even moments allowed us to assume the following general form of non-planar contribution to the anomalous dimension of twist-2 operators with arbitrary Lorentz spin:
\begin{eqnarray}
\gamma_{uni,\,np}\,(j)&\ =&\
-640\; S_1^2(j-2)\,\frac{12}{N_c^2}\ \zf\,g^8\,+\,...,\qquad S_1(j)=\sum_{i=1}^j\frac{1}{i}\,.
\label{resnpuad}
\end{eqnarray}
However, this result is in contradiction with the usual large-$j$ behavior of anomalous dimension, which is expected to be proportional  $\ln j$ (see Ref.\cite{Alday:2007mf}), while from Eq.(\ref{resnpuad}) we obtain $(\ln j)^2$.

In this paper we present the result of calculations of non-planar contribution for the first three even moments of the four-loop anomalous dimension of twist-2 operators in $\cN=4$ SYM theory, which completes our result Eq.(\ref{resnpK}).
Moreover, during these calculations we have obtained the first even moment of the non-planar contribution to the non-singlet anomalous dimension of Wilson twist-2 operator at fourth order in perturbative QCD.
The result for the first even moment of the four-loop non-singlet anomalous dimension of Wilson twist-2 operator in QCD can be found in Ref.\cite{Baikov:2006ai}, but our result for the non-planar contribution contains full color and flavor structures and the calculations are performed with a different method\footnote{Note, that there is all-loop prediction for the ${\mathcal O}(1/N_f)$ contribution to the non-singlet anomalous dimension of twist-2 operators in QCD~\cite{Gracey:1994nn}.}.

The calculations were performed in the same way, as in our previous work~\cite{Velizhanin:2008pc}.
We consider the following ``QCD-like'' colour and $SU(4)$ singlet local Wilson twist-2 operators:
\begin{eqnarray}
\mathcal{O}_{\mu _{1},...,\mu _{\M}}^{g} &=&\hat{S}
G_{\rho \mu_{1}}^{a}{\mathcal D}_{\mu _{2}}
{\mathcal D}_{\mu _{3}}...{\mathcal D}_{\mu _{\M-1}}G_{\rho \mu _{\M}}^a\,,
\label{ggs}\\
\mathcal{O}_{\mu _{1},...,\mu _{\M}}^{\lambda } &=&\hat{S}
\bar{\lambda}_{i}^{a}\gamma _{\mu _{1}}
{\mathcal D}_{\mu _{2}}...{\mathcal D}_{\mu _{\M}}\lambda ^{a\;i}\,, \label{qqs}\\
\mathcal{O}_{\mu _{1},...,\mu _{\M}}^{\phi } &=&\hat{S}
\bar{\phi}_{r}^{a}{\mathcal D}_{\mu _{1}}
{\mathcal D}_{\mu _{2}}...{\mathcal D}_{\mu _{\M}}\phi _{r}^{a}\,,\label{phphs}
\end{eqnarray}
where ${\mathcal D}_{\mu_i }$ are covariant derivatives.
The spinors $\lambda _{i}$ and field tensor $G_{\rho \mu }$ describe gauginos and gauge fields, respectively, and $\phi _{r}$ are the complex scalar fields appearing in the ${\mathcal N}=4$ SYM theory.
Indices $i=1,\cdots ,4$ and $r=1,\cdots ,3$ refer to $SU(4)$ and $SO(6)\simeq SU(4)$ groups of inner symmetry, respectively.
The symbol $\hat{S}$ implies a symmetrization of each tensor in the Lorentz indices $\mu_{1},...,\mu _{\M}$ and a subtraction of its traces.
These operators mix with each other under renormalization and the eigenvalues of the matrix are expressed through the universal anomalous dimension
(see our papers in Ref.\cite{Anselmi:1998ms} for details)
\begin{equation}
\gamma_{uni}(\M)=\sum_{n=0}^{\infty}\gamma_{uni}^{(n)}(\M)\,g^{2(n+1)}\,.
\end{equation}

We will be interested in the following leading order multiplicative renormalizable combinations of the operators (\ref{ggs})-(\ref{phphs}) with $j\!=\!2$ (see details in Ref.\cite{Velizhanin:2008pc})\footnote{The coefficients in the front of the operators $\Op^g_{\mu\nu}$, $\Op^\lambda_{\mu\nu}$ and $\Op^\phi_{\mu\nu}$ in Eqs.(\ref{mrop1j2})-(\ref{mrop3j2}) are the same (up to common factor) as in the conformal operators $\Xi_{\mu\nu}$, $\Sigma_{\mu\nu}$ and $T_{\mu\nu}$ from the first paper in Ref.\cite{Anselmi:1998ms}.}:
\begin{eqnarray}
\Op^T_{\mu\nu} & = &
\Op^g_{\mu\nu}
+ \Op^\lambda_{\mu\nu}
+ \Op^\phi_{\mu\nu}\,,
\label{mrop1j2}\\[1mm]
\Op^\Sigma_{\mu\nu} & = &
- 2\,\Op^g_{\mu\nu}
+ \Op^\lambda_{\mu\nu}
+ 2\,\Op^\phi_{\mu\nu}\,,
\label{mrop2j2}\\[1mm]
\Op^\Xi_{\mu\nu} & = &
- \frac{1}{4}\Op^g_{\mu\nu}
+ \Op^\lambda_{\mu\nu}
- \frac{3}{2}\Op^\phi_{\mu\nu}\,.
\label{mrop3j2}
\end{eqnarray}
Operator $\Op^T_{\mu\nu}$ is the stress tensor. Its anomalous dimension is equal to zero and corresponds to $\gamma^{(0)}_{uni}(j\!=\!2)$.
Operator $\Op^\Sigma_{\mu\nu}$ has the same anomalous dimension as the Konishi operator, which corresponds to $\gamma^{(0)}_{uni}(j\!=\!4)$ and we know  its anomalous dimension (see Eq.(\ref{resnpK})).
Operator $\Op^\Xi_{\mu\nu}$ has the anomalous dimension, which corresponds to the value of universal anomalous dimension $\gamma^{(0)}_{uni}(j)$ with $j\!=\!6$.

In the next-to-leading order operators~(\ref{mrop1j2})-(\ref{mrop3j2}) mix with each other under renormalization.
It is related with the breaking of the conformal invariance if we consider more general conformal operators~\cite{Makeenko:1980bh}.
Breaking of conformal invariance is controlled by the conformal Ward identity~\cite{Mueller:1991gd} (see also Ref.\cite{Mikhailov:1985cm}), which allows obtain the results for the anomalous dimensions of the conformal operators in $\ell^{th}$-loops order with additional $(\ell\!-\!1)$-loops calculations~\cite{Belitsky:1998vj}.

If we go to the four loops we can do the same as in the leading order for the contribution to the universal anomalous dimension, which appears for the first time at this order because there are no additional contributions either from the renormalizations or from the conformal Ward identities.
We used this property for the calculation of $\zf$ contribution and will use here for the calculation of the non-planar contribution.

Thus, we need to calculate the matrix elements for the operators $\Op^T_{\mu\nu}$, $\Op^\Sigma_{\mu\nu}$ and $\Op^\Xi_{\mu\nu}$ sandwiched between fermion states (as the most simple case) and look only at the pole with quartic Casimir operator $d_{44} =N^2(N^2+36)/24$.
This can be done with our program BAMBA 
based on the algorithm of Laporta~\cite{Laporta:2001dd} (see also Ref.\cite{Misiak:1994zw,Czakon:2004bu}), which we used in our previous calculations.

All calculations were performed with FORM~\cite{Vermaseren:2000nd}, using FORM package COLOR~\cite{vanRitbergen:1998pn} for evaluation of the color traces and with the Feynman rules from Refs.\cite{Gliozzi:1976qd}.
For the dealing with a huge number of diagrams we use a program DIANA~\cite{Tentyukov:1999is}, which call QGRAF~\cite{Nogueira:1991ex} to generate all diagrams.

In fact, we have computed the non-planar contributions to the four-loop anomalous dimensions $\gamma_{g\lambda}$, $\gamma_{\phi\lambda}$ and $\gamma_{\lambda\lambda}$ of the operators $\Op^g_{\mu\nu}$, $\Op^\lambda_{\mu\nu}$ and $\Op^\phi_{\mu\nu}$ sandwiched between the fermion states to have a possibility to combine them with the coefficients from Eqs.(\ref{mrop1j2})-(\ref{mrop3j2}) for the additional check.
We have obtained the following results for the non-planar contributions to the four-loop anomalous dimensions of the operators $\Op^T_{\mu\nu}$, $\Op^\Sigma_{\mu\nu}$ and $\Op^\Xi_{\mu\nu}$ sandwiched between fermion states:
\begin{eqnarray}
\Gamma^{np}_{T_{\mu\nu}} & = &
0\,,
\label{mrop1j2ad}\\[5mm]
\Gamma^{np}_{\Sigma_{\mu\nu}} & = &
- 360\ \zf\, \frac{48\,g^8}{N_c^2}+\mathcal{O}\!\left(g^{10}\right)\,,
\label{mrop2j2ad}\\[2mm]
\Gamma^{np}_{\Xi_{\mu\nu}} & = &
\frac{25}{9}\bigg(21 + 70 \ \zt - 250 \ \zf \bigg)\,  \frac{48\,g^8}{N_c^2}+\mathcal{O}\!\left(g^{10}\right)\,.
\label{mrop3j2ad}
\end{eqnarray}
The first two results coincide with the known results~\cite{Velizhanin:2009gv}.
The $\zf$ part of the third result we already know~\cite{Velizhanin:2009gv} and we suggest the following general expression for the $\zf$ contribution to the non-planar four-loop universal anomalous dimension (see Ref.\cite{Velizhanin:2009gv}):
\begin{equation}\label{HSZ5ResS1}
\gamma_{uni,np,\,\zf}^{(3)}(\M)\ =\ -\,160\,\HS_{1}^2(\M-2)
\end{equation}
with
\begin{equation}\label{HSZ5ResALL}
\gamma_{uni,np}^{(3)}(\M)=\left(\gamma_{uni,np,\,\zf}^{(3)}(\M)\,\zf+\gamma_{uni,np,\,\zt}^{(3)}(\M)\,\zt+\gamma_{uni,np,rational}^{(3)}(\M)\right)\frac{48}{N_c^2}\,.
\end{equation}
Now we can try to reconstruct a general form of the  $\zt$ part of the non-planar contribution to the four-loop universal anomalous dimension $\gamma_{uni,\,np}^{(3)}$.
Following to the principle of maximal transcendentality~\cite{Kotikov:2002ab}, in the fourth order of the perturbative theory the transcendentality level of the universal anomalous dimension is equal to $7$ and the transcendentality of $\zt$ is equal to $3$.
Let's suppose, that the reciprocity~\cite{Dokshitzer:2005bf} will hold also for the non-planar contribution.
Thus, the basis for ansatz will consist of the binomial harmonic sums, defined through (see Ref.\cite{Vermaseren:1998uu} and our papers in Ref.\cite{Bajnok:2008bm} for details)
\begin{equation}
\HBS_{i_1,\ldots,i_k}(N)=(-1)^N\sum_{j=1}^{N}(-1)^j\binom{N}{j}\binom{N+j}{j}\HS_{i_1,...,i_k}(j)\,,
\end{equation}
which should have transcendentality $4$.
In the above equation $\HS_{i_1, \ldots ,i_k}$ are the harmonic sums~\cite{Vermaseren:1998uu}
\begin{equation} \label{vhs}
S_a (N)=\sum^{N}_{j=1} \frac{(\mbox{sgn}(a))^{j}}{j^{\vert a\vert}}\, , \qquad
S_{a_1,\ldots,a_n}(N)=\sum^{N}_{j=1} \frac{(\mbox{sgn}(a_1))^{j}}{j^{\vert a_1\vert}}
\,S_{a_2,\ldots,a_n}(j)
\end{equation}
and the indices $i_1,\ldots,i_k$ are \textit{positive}.
There are $2^3=8$ such binomial harmonic sums:
\begin{equation}
\HBS_4,\,\HBS_{3,1},\,\HBS_{2,2},\,\HBS_{2,1,1},\,\HBS_{1,3},\,\HBS_{1,2,1},\,\HBS_{1,1,2},\,\HBS_{1,1,1,1}\,.
\end{equation}
The basis from above binomial sums can be rewritten in the following equivalent form:
\begin{equation}\label{ansatzHBS1}
\HBS_4,\,\HBS_{3,1},\,\HBS_{2,2},\,\HBS_{2,1,1},\,\HBS_{1}\, \HBS_{3},\,\HBS_{1}\, \HBS_{2,1},\,\HBS_{1}^2\, \HBS_{2},\,\HBS_{1}^4\,.
\end{equation}
The common factor in $\Gamma_{\Xi_{\mu\nu}} $ Eq.(\ref{mrop3j2ad}) can be written as $4\,\HS_1(4)=2\,\HBS_1(4)$.
Therefore, let's try ansatz, which consists of the binomial harmonic sums in Eq.(\ref{ansatzHBS1}) with $\HBS_1$ except for the last sum:
\begin{equation}
\HBS_{1}\,\HBS_{3},\,\HBS_{1}\,\HBS_{2,1},\,\HBS_{1}^2\,\HBS_{2}\,.
\end{equation}
Note, that this basis contains the same binomial sums as the $\zt$ part of the planar contribution to the four-loop universal anomalous dimension
\begin{equation}
\gamma_{uni,pl,\,\zt}^{(3)}(\M)=
64\,\HBS_{1}\big(\HBS_{3}-\HBS_{2,1}\big)+64\,\HBS_{1}^2\,\HBS_{2}\,,
\end{equation}
where the first term comes from the dressing phase while the second term comes from the wrapping corrections. Here and in the following the argument of the (binomial) harmonic sums is $j\!-\!2$.
Really, we have only two non-trivial values (\ref{mrop2j2ad}) and  (\ref{mrop3j2ad}), then the solution has one free integer parameter $x$
\begin{equation}
\gamma_{uni,np,\,\zt}^{(3)}(\M)=\HBS_{1}\big(9\,(24-x)\,\HBS_{3}+(x-72)\,\HBS_{2,1}+x\,\HBS_1\,\HBS_2\big)\,.
\end{equation}
In principal, we can fix $x$ with some reasonable conditions, for example:\\
1) putting the coefficient of $\HBS_1\,\HBS_3$ equal to zero
\begin{equation}
\gamma_{uni,np,\,\zt}^{(3)}(\M)=24\,\HBS_{1}\big(\HBS_1\,\HBS_2-2\,\HBS_{2,1}\big)\,;\label{hbs1hbs3}
\end{equation}
2) putting the coefficient of $\HBS_1\,\HBS_{2,1}$ equal to zero
\begin{equation}
\gamma_{uni,np,\,\zt}^{(3)}(\M)=72\,\HBS_{1}\big(\HBS_1\,\HBS_2-6\,\HBS_{3}\big)\,;\label{hbs1hbs21}
\end{equation}
3) putting the coefficient of $\HBS_1^2\,\HBS_2$ equal to zero
\begin{equation}
\gamma_{uni,np,\,\zt}^{(3)}(\M)=72\,\HBS_{1}\big(3\,\HBS_3-\HBS_{2,1}\big)\,;\label{hbs12hbs3}
\end{equation}
4) split final expression into the part, which is similar to wrapping corrections in the planar limit, and the remnant
\begin{equation}
\gamma_{uni,np,\,\zt}^{(3)}(\M)=8\,\HBS_{1}\big(2\,\HBS_1\,\HBS_2+9\,\HBS_{3}-7\,\HBS_{2,1}\big)\,.\label{hbsAhbsA}
\end{equation}
It is interesting that Eq.(\ref{hbs1hbs3}) gives zero at  $j=3$ as well as for $j=2$ and $j=4$.
So, we believe, that Eq.(\ref{hbs1hbs3}) is the preferable expression for the $\zt$ contribution to the non-planar four-loop universal anomalous dimension, which can be rewritten in the terms of the usual harmonic sums as
\begin{equation}
\gamma_{uni,np,\,\zt}^{(3)}(\M)=-192\,\HS_{1}\big(\HS_1\,\HS_{-2}+\,\HS_{3}\big)\,.\label{hs1hs3}
\end{equation}

As we wrote in the beginning, in the course of calculations we can easily obtain the non-planar contribution to the first even moment of the four-loop non-singlet anomalous dimension of Wilson twist-2 operators in QCD.
Formally, we should consider the following Wilson twist-2 operator (cf. Eq.(\ref{qqs}))
\begin{equation}
\mathcal{O}_{\mu _{1},...,\mu _{\M}}^{i} =\hat{S}
\bar{\psi}\Lambda^i\gamma _{\mu _{1}}
{\mathcal D}_{\mu _{2}}...{\mathcal D}_{\mu _{\M}}\psi\,,\qquad i=1,...,8\,, \label{qqQCD}
\end{equation}
where $\psi$ is the quark field and $\Lambda^i$ is the flavour group generator of $SU(N_F)$.
This operator is multiplicative renormalizable, that is does not mix with all other Wilson twist-2 operators in QCD.
To find its anomalous dimension we should calculate the matrix element of this operator sandwiched between quark states with the same flavour.
For the first even moment ($j=2$) of operator $\mathcal{O}_{\mu _{1},...,\mu _{\M}}^{i}$ from Eq.(\ref{qqQCD})
\begin{equation}
\mathcal{O}_{\mu _{1},\mu _{2}} =\hat{S}
\bar{\psi}\Lambda^i\gamma _{\mu _{1}}
{\mathcal D}_{\mu _{2}}\psi \label{qqQCDj2}
\end{equation}
all necessary diagrams are included in our above calculations.
But we should remember that quarks are in the fundamental representation of color group, while gauginos in $\cN=4$ SYM are in the adjoint representation.
This leads to the appearance of additional quartic Casimir operators of the fundamental and adjoint representations (see Refs.\cite{vanRitbergen:1998pn} and~\cite{Czakon:2004bu})
\begin{eqnarray}
\frac{d_F^{abcd}d_F^{abcd}}{N_A}&\ =\ &\frac{N_c^4-6N_c^2+18}{96N_c^2}\,,\\
\frac{d_F^{abcd}d_A^{abcd}}{N_A}&\ =\ &\frac{N_c(N_c^2+6)}{8}\,,\\
\frac{d_A^{abcd}d_A^{abcd}}{N_A}&\ =\ &d_{44}\ =\ \frac{N_c^2(N_c^2+36)}{24}
\end{eqnarray}
and for the color group $SU(N_c)$ the more simple Casimir operators are:
\begin{equation}
T_F=\frac12\,,\qquad
C_F=\frac{N_c^2-1}{2N_c}\,,\qquad
C_A=N_c\,,\qquad
N_A=N_c^2-1\,.
\end{equation}
So, our result for the non-planar contribution to the four-loop non-singlet anomalous dimension of Wilson twist-2 operator (\ref{qqQCDj2}) is given by
\begin{equation}
\gamma^{(3)}_{ns,np}(j\!=\!2)=
32\,\Big(13 + 16 \zt - 40 \zf\Big)n_f\frac{d_F^{abcd}d_F^{abcd}}{N_F}
-8\,\Big(23 - 62 \zt - 160 \zf\Big)\frac{d_F^{abcd}d_A^{abcd}}{N_F}
\,, \label{gamma3nsnp}
\end{equation}
where $n_f$ is the number of active quarks and
\begin{equation}
\gamma_{ns}=\sum_{n=0}^\infty\gamma^{(n)}_{ns}a_s^{(n+1)}\,,\qquad a_s=\frac{\alpha_s}{4\pi}\,.
\end{equation}
We are going to complete this result with the rest parts in the future calculations.
Unfortunately, the obtained result can not be compared with the existing result for the first even moment of non-singlet anomalous dimension from~\cite{Baikov:2006ai}, as in the result from Ref.\cite{Baikov:2006ai} all  Casimir operators are written explicitly for QCD with three active quarks (i.e. for the gauge group $SU(3)$ with $n_f=3$), so it is impossible to separate non-planar contribution from planar.
Comparison with the result from Ref.\cite{Gracey:1994nn} is impossible, because this result is proportional to $(n_f)^{i-1}a_s^i$, that is the number of possible quark loops, which is equal to three at four-loop order.

To conclude, we note that our guess about general form of $\zt$ contribution to the non-planar four-loop universal anomalous dimension of twist-2 operators in $\cN=4$ SYM theory (\ref{hs1hs3}) will be checked by calculations of the next even moments in our forthcoming calculations.
We hope, that these new calculations together with the different constraints will give enough information for the reconstruction of the rational part to extract the non-planar scaling function.
In any case our result~(\ref{mrop3j2ad}) of the {\it full direct} calculations for the non-planar contribution to the four-loop anomalous dimension of twist-2 operator is new and can be used in the investigations of non-planar aspects of AdS/CFT-correspondence.

 \subsection*{Acknowledgments}
We would like to thank K.G. Chetyrkin, L.N. Lipatov, A.I. Onishchenko, A.V. Smirnov, V.A. Smirnov and M. Staudacher
for useful discussions.
This work is supported by RFBR grants 10-02-01338-a, RSGSS-65751.2010.2.
We thank the Max Planck Institute for Gravitational Physics at Potsdam for hospitality while working on parts of this project.


\begin{thebibliography}{00}


\bibitem{Gross:1973ju}
  D.~J.~Gross and F.~Wilczek,
  Phys.\ Rev.\  D {\bf 8} (1973) 3633;
  H.~Georgi and H.~D.~Politzer,
  Phys.\ Rev.\  D {\bf 9} (1974) 416;
  E.~G.~Floratos, D.~A.~Ross and C.~T.~Sachrajda,
  Nucl.\ Phys.\  B {\bf 129} (1977) 66;
  E.~G.~Floratos, D.~A.~Ross and C.~T.~Sachrajda,
  Nucl.\ Phys.\  B {\bf 152} (1979) 493;
  A.~Gonzalez-Arroyo, C.~Lopez and F.~J.~Yndurain,
  Nucl.\ Phys.\  B {\bf 153} (1979) 161;
  A.~Gonzalez-Arroyo and C.~Lopez,
  Nucl.\ Phys.\  B {\bf 166} (1980) 429;
  R.~Hamberg and W.~L.~van Neerven,
  Nucl.\ Phys.\  B {\bf 379} (1992) 143;
  R.~K.~Ellis and W.~Vogelsang,
  arXiv:hep-ph/9602356;
  S.~Moch, J.~A.~M.~Vermaseren and A.~Vogt,
  Nucl.\ Phys.\  B {\bf 688} (2004) 101;
  A.~Vogt, S.~Moch and J.~A.~M.~Vermaseren,
  Nucl.\ Phys.\  B {\bf 691} (2004) 129.

\bibitem{Vermaseren:2005qc}
  J.~A.~M.~Vermaseren, A.~Vogt and S.~Moch,
  Nucl.\ Phys.\  B {\bf 724} (2005) 3.

\bibitem{Maldacena:1997re}
  J.~M.~Maldacena,
  Adv.\ Theor.\ Math.\ Phys.\  {\bf 2} (1998) 231;
%
  S.~S.~Gubser, I.~R.~Klebanov and A.~M.~Polyakov,
  Phys.\ Lett.\  B {\bf 428} (1998) 105;
%
  E.~Witten,
  Adv.\ Theor.\ Math.\ Phys.\  {\bf 2} (1998) 253.

\bibitem{Anselmi:1998ms}
  D.~Anselmi,
  Nucl.\ Phys.\  B {\bf 541} (1999) 369;
L.~N.~Lipatov, in: \textit{Proc. of the Int. Workshop on very
high multiplicity physics}, Dubna, 2000, pp.159-176;
  M.~Bianchi, S.~Kovacs, G.~Rossi and Y.~S.~Stanev,
  Nucl.\ Phys.\  B {\bf 584} (2000) 216;
  G.~Arutyunov, B.~Eden, A.~C.~Petkou and E.~Sokatchev,
  Nucl.\ Phys.\  B {\bf 620} (2002) 380;
  F.~A.~Dolan and H.~Osborn,
  Nucl.\ Phys.\  B {\bf 629} (2002) 3;
  A.~V.~Kotikov, L.~N.~Lipatov and V.~N.~Velizhanin,
  Phys.\ Lett.\  B {\bf 557} (2003) 114;
  A.~V.~Kotikov, L.~N.~Lipatov, A.~I.~Onishchenko and V.~N.~Velizhanin,
  Phys.\ Lett.\  B {\bf 595} (2004) 521;
  F.~Fiamberti, A.~Santambrogio, C.~Sieg and D.~Zanon,
  Phys.\ Lett.\  B {\bf 666} (2008) 100;
  F.~Fiamberti, A.~Santambrogio, C.~Sieg and D.~Zanon,
  Nucl.\ Phys.\  B {\bf 805} (2008) 231;
  V.~N.~Velizhanin,
  JETP Lett.\  {\bf 89} (2009) 6;
  V.~N.~Velizhanin,
  Phys.\ Lett.\  B {\bf 676} (2009) 112;
  F.~Fiamberti, A.~Santambrogio and C.~Sieg,
  JHEP {\bf 1003} (2010) 103.

\bibitem{Bajnok:2008bm}
  Z.~Bajnok and R.~A.~Janik,
  Nucl.\ Phys.\  B {\bf 807} (2009) 625;
  A.~V.~Kotikov, L.~N.~Lipatov, A.~Rej, M.~Staudacher and V.~N.~Velizhanin,
  J.\ Stat.\ Mech.\  {\bf 0710} (2007) P10003;
  B.~Eden, C.~Jarczak and E.~Sokatchev,
  Nucl.\ Phys.\  B {\bf 712} (2005) 157;
  Z.~Bajnok, R.~A.~Janik and T.~Lukowski,
  Nucl.\ Phys.\  B {\bf 816} (2009) 376;
  M.~Beccaria, V.~Forini, T.~Lukowski and S.~Zieme,
  JHEP {\bf 0903} (2009) 129;
  Z.~Bajnok, A.~Hegedus, R.~A.~Janik and T.~Lukowski,
  Nucl.\ Phys.\  B {\bf 827} (2010) 426;
  T.~Lukowski, A.~Rej and V.~N.~Velizhanin,
  Nucl.\ Phys.\  B {\bf 831} (2010) 105;
  V.~N.~Velizhanin,
  JHEP {\bf 1011} (2010) 129.

\bibitem{Minahan:2002ve}
  J.~A.~Minahan and K.~Zarembo,
  JHEP {\bf 0303} (2003) 013;
%
  N.~Beisert, C.~Kristjansen and M.~Staudacher,
  Nucl.\ Phys.\  B {\bf 664} (2003) 131;
%
  N.~Beisert and M.~Staudacher,
  Nucl.\ Phys.\  B {\bf 670} (2003) 439;
%
  N.~Beisert, V.~Dippel and M.~Staudacher,
  JHEP {\bf 0407} (2004) 075;
%
  M.~Staudacher,
  JHEP {\bf 0505} (2005) 054;
%
  N.~Beisert and M.~Staudacher,
  Nucl.\ Phys.\  B {\bf 727} (2005) 1;
%
  Z.~Bern, M.~Czakon, L.~J.~Dixon, D.~A.~Kosower and V.~A.~Smirnov,
  Phys.\ Rev.\  D {\bf 75} (2007) 085010;
  N.~Beisert, B.~Eden and M.~Staudacher,
  J.\ Stat.\ Mech.\  {\bf 0701} (2007) P021;
  N.~Beisert, T.~McLoughlin and R.~Roiban,
  Phys.\ Rev.\  D {\bf 76} (2007) 046002.
  I.~Bena, J.~Polchinski and R.~Roiban,
  Phys.\ Rev.\  D {\bf 69} (2004) 046002;
%
  L.~Dolan, C.~R.~Nappi and E.~Witten,
  JHEP {\bf 0310} (2003) 017;
%
  V.~A.~Kazakov, A.~Marshakov, J.~A.~Minahan and K.~Zarembo,
  JHEP {\bf 0405} (2004) 024;
%
  N.~Beisert, V.~A.~Kazakov, K.~Sakai and K.~Zarembo,
  Commun.\ Math.\ Phys.\  {\bf 263} (2006) 659;
%
  N.~Beisert, V.~A.~Kazakov, K.~Sakai and K.~Zarembo,
  JHEP {\bf 0507} (2005) 030;
%
  G.~Arutyunov, S.~Frolov and M.~Staudacher,
  JHEP {\bf 0410} (2004) 016;
%
  N.~Beisert and A.~A.~Tseytlin,
  Phys.\ Lett.\  B {\bf 629} (2005) 102.

\bibitem{Arutyunov:2009zu}
  G.~Arutyunov and S.~Frolov,
  JHEP {\bf 0903} (2009) 152;
%
  N.~Gromov, V.~Kazakov and P.~Vieira,
  Phys.\ Rev.\ Lett.\  {\bf 103} (2009) 131601;
%
  D.~Bombardelli, D.~Fioravanti and R.~Tateo,
  J.\ Phys.\ A  {\bf 42} (2009) 375401;
%
  N.~Gromov, V.~Kazakov, A.~Kozak and P.~Vieira,
  Lett.\ Math.\ Phys.\  {\bf 91} (2010) 265;
%
  G.~Arutyunov and S.~Frolov,
  JHEP {\bf 0905} (2009) 068;
%
  G.~Arutyunov, S.~Frolov and R.~Suzuki,
  JHEP {\bf 1005} (2010) 031.

\bibitem{Luscher:1985dn}
  M.~Luscher,
  Commun.\ Math.\ Phys.\  {\bf 104} (1986) 177;
%
  M.~Luscher,
  Commun.\ Math.\ Phys.\  {\bf 105} (1986) 153.

\bibitem{Velizhanin:2009gv}
  V.~N.~Velizhanin,
  JETP Lett.\  {\bf 89} (2009) 593.

\bibitem{Alday:2007mf}
  L.~F.~Alday and J.~M.~Maldacena,
  JHEP {\bf 0711} (2007) 019.

\bibitem{Baikov:2006ai}
  P.~A.~Baikov and K.~G.~Chetyrkin,
  Nucl.\ Phys.\ Proc.\ Suppl.\  {\bf 160} (2006) 76.

\bibitem{Gracey:1994nn}
  J.~A.~Gracey,
  Phys.\ Lett.\  B {\bf 322} (1994) 141.

\bibitem{Velizhanin:2008pc}
  V.~N.~Velizhanin,
  Phys.\ Lett.\  B {\bf 676} (2009) 112.

\bibitem{Makeenko:1980bh}
  Yu.~M.~Makeenko,
  Sov.\ J.\ Nucl.\ Phys.\  {\bf 33} (1981) 440;
  T.~Ohrndorf,
  Nucl.\ Phys.\  B {\bf 198} (1982) 26;
  A.~P.~Bukhvostov, G.~V.~Frolov, L.~N.~Lipatov and E.~A.~Kuraev,
  Nucl.\ Phys.\  B {\bf 258} (1985) 601;
  A.~I.~Onishchenko and V.~N.~Velizhanin,
  arXiv:hep-ph/0309222;
  A.~V.~Belitsky, S.~E.~Derkachov, G.~P.~Korchemsky and A.~N.~Manashov,
  Phys.\ Rev.\  D {\bf 70} (2004) 045021.


\bibitem{Mueller:1991gd}
  D.~Mueller,
  Z.\ Phys.\  C {\bf 49} (1991) 293;
  D.~Mueller,
  Phys.\ Rev.\  D {\bf 49} (1994) 2525.

\bibitem{Mikhailov:1985cm}
  S.~V.~Mikhailov and A.~V.~Radyushkin,
  Nucl.\ Phys.\  B {\bf 273} (1986) 297.

\bibitem{Belitsky:1998vj}
  A.~V.~Belitsky and D.~Mueller,
  Nucl.\ Phys.\  B {\bf 527} (1998) 207;
  A.~V.~Belitsky and D.~Mueller,
  Nucl.\ Phys.\  B {\bf 537} (1999) 397;
  A.~V.~Belitsky, D.~Mueller and A.~Schafer,
  Phys.\ Lett.\  B {\bf 450} (1999) 126;
  A.~V.~Belitsky and D.~Mueller,
  Phys.\ Rev.\  D {\bf 65} (2002) 054037;
  A.~I.~Onishchenko and V.~N.~Velizhanin,
  JHEP {\bf 0402} (2004) 036.


\bibitem{Laporta:2001dd}
  S.~Laporta,
  Int.\ J.\ Mod.\ Phys.\  A {\bf 15} (2000) 5087.

\bibitem{Misiak:1994zw}
  M.~Misiak and M.~Munz,
  Phys.\ Lett.\  B {\bf 344} (1995) 308;
%
  K.~G.~Chetyrkin, M.~Misiak and M.~Munz,
  Nucl.\ Phys.\  B {\bf 518} (1998) 473.

\bibitem{Czakon:2004bu}
  M.~Czakon,
  Nucl.\ Phys.\  B {\bf 710} (2005) 485.

\bibitem{Vermaseren:2000nd}
  J.~A.~M.~Vermaseren,
  arXiv:math-ph/0010025.

\bibitem{vanRitbergen:1998pn}
  T.~van Ritbergen, A.~N.~Schellekens and J.~A.~M.~Vermaseren,
  Int.\ J.\ Mod.\ Phys.\  A {\bf 14} (1999) 41.

\bibitem{Gliozzi:1976qd}
  F.~Gliozzi, J.~Scherk and D.~I.~Olive,
  Nucl.\ Phys.\  B {\bf 122} (1977) 253;
  L.~V.~Avdeev, O.~V.~Tarasov and A.~A.~Vladimirov,
  Phys.\ Lett.\  B {\bf 96} (1980) 94.

\bibitem{Tentyukov:1999is}
  M.~Tentyukov and J.~Fleischer,
  Comput.\ Phys.\ Commun.\  {\bf 132} (2000) 124.

\bibitem{Nogueira:1991ex}
  P.~Nogueira,
  J.\ Comput.\ Phys.\  {\bf 105} (1993) 279.

\bibitem{Kotikov:2002ab}
  A.~V.~Kotikov and L.~N.~Lipatov,
  Nucl.\ Phys.\  B {\bf 661} (2003) 19.

\bibitem{Dokshitzer:2005bf}
  Yu.~L.~Dokshitzer, G.~Marchesini and G.~P.~Salam,
  Phys.\ Lett.\  B {\bf 634} (2006) 504;
  Yu.~L.~Dokshitzer and G.~Marchesini,
  Phys.\ Lett.\  B {\bf 646} (2007) 189.

\bibitem{Vermaseren:1998uu}
  J.~A.~M.~Vermaseren,
  Int.\ J.\ Mod.\ Phys.\  A {\bf 14} (1999) 2037.

\end{thebibliography}
\end{document}